\documentstyle[graphicx]{mn}

\newif\ifAMStwofonts
\AMStwofontstrue

\def\pg{{PG0844+349}}

\def\msun{{\rm M_{\odot}}}

\def\rp{{R_{\rm ph}}}
\def\rs{{R_{\rm s}}}
\def\mo{{\dot M_{\rm out}}}
\def\me{{\dot M_{\rm Edd}}}

\def\einstein{{\it Einstein}}
\def\exosat{{\it EXOSAT}}
\def\xmm{{\it XMM-Newton}}
\def\chandra{{\it Chandra}}

\def\et{{et al.\ }}

\def\asca{{\it ASCA}}

\newcommand{\ls}{\mathrel{\hbox{\rlap{\hbox{\lower4pt\hbox{$\sim$}}}\hbox{$<$}}}}
\newcommand{\gs}{\mathrel{\hbox{\rlap{\hbox{\lower4pt\hbox{$\sim$}}}\hbox{$>$}}}}

\def\arcs{{\hbox{$^{\prime\prime}$}}}

\def\Msun{\hbox{$\rm ~M_{\odot}$}}

\def\H0{{\rm ~km~s^{-1}~Mpc^{-1}}}

\def\msun{M_{\rm \odot}}

\def\et{{et al.}}

\title[A massive ionised outflow]
        {Evidence of a high velocity ionised outflow in a second narrow line quasar \pg.}
\author[K.A.Pounds \et]
        {K.A.Pounds,
	A.R.King,
	K.L.Page, 
	P.T.O'Brien\\
	Department of Physics and Astronomy, University of Leicester,
Leicester, LE1 7RH, UK\\}

\date{Accepted 21 August 2003; Submitted 29 May 2003; Revised 13 August 2003}
\pagerange{\pageref{firstpage}--\pageref{lastpage}}
\pubyear{2003}
\begin{document}
\maketitle
\label{firstpage}

\begin{abstract}

Following the discovery of X-ray absorption in a high velocity outflow from the bright quasar PG1211+143 we have searched
for  similar features in \xmm\ archival data of a second (high accretion rate) quasar \pg. Evidence is found for several faint
absorption lines in both the EPIC and RGS spectra, whose most likely identification with resonance transitions in H-like Fe, S, and Ne
implies an origin in highly ionised matter with an outflow velocity of order $\sim$0.2c. The  line equivalent widths require
a line-of-sight column density of $N_{H}$$\sim$$4\times10^{23}\rm{cm}^{-2}$, at an ionisation parameter of
log$\xi$$\sim$3.7.  Assuming a radial outflow being driven by radiation pressure from the inner accretion disc, as suggested
previously for PG1211+143, the flow in \pg\ is also likely to be optically thick, in this case within $\sim$25 Schwarzschild radii.
If confirmed by better data our analysis suggests that a high velocity, highly ionised outflow is likely to be a significant component in the mass and energy
budgets of AGN accreting at or above the Eddington rate.

\end{abstract}


\section{Introduction}

The additional sensitivity of \xmm\ and \chandra, particularly in the energy band above the Fe K edge at
$\sim$7 keV, has recently provided the first X-ray evidence for high velocity outflows from quasars (Chartas \et\ 2002,
Pounds \et\ 2003, Reeves \et\ 2003). The
deduced ionisation parameters and column densities are substantially higher than for the optical/UV lines
seen in Broad Absorption Line (BAL) quasars, with at least three important consequences. First, the detection of a
high velocity outflow in PG 1211+143 (Pounds \et\ 2003), coupled with that reported here for \pg, imply that
this could be a property common to many bright quasars. Second, interpreting the observed absorption as
arising in a radial outflow leads inevitably to the prediction that the flow will be
optically thick close to the black hole. Third, the outflow is likely to be a significant component in the mass and
energy budgets of AGN accreting at or above the Eddington rate. 

In this paper we report on the spectral analysis of a $\sim$25 ksec \xmm\ observation of the bright quasar
\pg\ taken from the \xmm\ data archive. We assess the results in terms of a recent model describing the characteristics
of outflows from black holes accreting at or above the Eddington rate (King and Pounds 2003).

\pg\ is a low redshift ($z=0.064$), optically bright quasar, with strong Fe II emission and relatively narrow permitted
optical  emission lines (Boroson and Green 1992, Kaspi 2000). It was detected as a strong soft X-ray source in the ROSAT
sky survey, but had faded by a factor of 6 in a pointed observation 6 months later (Rachen \et\ 1996), when it could be
described as `X-ray quiet' by comparison with its optical flux (Yuan \et\ 1998). Comparison with earlier soft X-ray
detections with \einstein\ (Kriss 1988) and \exosat\ (Malaguti \et\ 1994) suggested the `X-ray quiet' phase is relatively
rare. A later \asca\  observation did indeed find \pg\ to be again in a bright state, with a 2--10 keV power law index
$\Gamma$$\sim$1.98 and evidence for Fe K emission with equivalent width EW$\sim$300~eV (Wang \et\ 2000). \pg\ has a
typical X-ray luminosity (2--10~keV) of $5\times$$10^{43}$~erg s$^{-1}$, for $ H_0 = 75 $~km\,s$^{-1}$\,Mpc$^{-1}$, and
lies behind a low Galactic column of $N_{H}=3.32\times10^{20}\rm{cm}^{-2}$ (Murphy \et\ 1996).

\section{Observation and data reduction}

\pg\ was observed by \xmm\ on 2000 November 5 for $\sim$25 ksec. An analysis of the  EPIC spectrum (Brinkmann \et\ 2003) reported a
strong soft X-ray excess, which the authors discuss in terms of Comptonisation of thermal accretion disc photons. Given the recent
detection of `blue-shifted' absorption lines in a very similar PG quasar (Pounds \et\ 2003), we have re-examined the \xmm\
observation of \pg\ to search for further evidence of such features. We use the full \xmm\ X-ray data, including the EPIC pn camera
(Str\"{u}der \et 2001), which has the best sensitivity of any instrument flown to date in the $\sim$7-10 keV spectral band above
the Fe K absorption edge, the combined EPIC MOS cameras (Turner \et\ 2001), and the Reflection Grating Spectrometer/RGS (den Herder
\et\ 2001). The X-ray data were first screened with the latest XMM SAS v5.4 software and events corresponding to patterns 0-4
(single and double pixel events) were selected for the pn data and patterns 0-12 for MOS1 and MOS2, the latter then being combined.
We considered using only pattern 0 data from the pn camera, since at a mean count rate of 7 s$^{-1}$  there was some evidence of
pile-up. However, given that our target was to search for narrow spectral features, we chose to retain the maximum high energy
count rate at the expense of some (minor) loss of energy resolution. A low energy cut of 300 eV was applied to all X-ray data and
known hot or bad pixels were removed. We extracted source counts within a circular region of 45\arcs\ radius defined around the
centroid position of \pg, with the background being taken from a similar region, offset from but close to the source. Particular
care was taken to locate the background region for the pn camera well inside the `shadow' caused by Cu K fluorescence in a circuit
board at the base of the camera. The nett exposures used for subsequent spectral analysis were 20.2 ks (pn), 46.7 ks (MOS1 + MOS2)
and 25 ks (each RGS). Individual spectra were binned to a minimum of 20 counts per bin, to facilitate use of the $\chi^2$
minimalisation technique in spectral fitting.  Spectral fitting was based on the Xspec package (Arnaud 1996) and used a grid of
ionised absorber models calculated with the XSTAR code (Kallman \et\ 1996).  All spectral fits include absorption due to the 
line-of-sight Galactic column of $N_{H}=3.32\times10^{20}\rm{cm}^{-2}$. Errors are quoted at the 90\% confidence level ($\Delta
\chi^{2}=2.7$ for one interesting parameter).

\section{2--11 keV spectrum}   

\subsection{Power law continuum} 

We began our analysis of \pg\ by confirming there were no obvious spectral changes with varying source flux and
then proceeded to fit the \xmm\ pn and MOS data integrated over the full $\sim$25 ksec observation. A simple power
law fit over the 2--11 keV band yielded a relatively steep photon index of $\Gamma$$\sim$2.13 (pn) and
$\Gamma$$\sim$2.08 (MOS), with a combined $\chi^{2}$/dof of 546/586. The statistical quality of this fit indicated
that any discrete spectral features were less strong than in the case of PG1211+143 (Pounds \et\ 2003). However,
visual examination of the data:powerlaw-model ratio (figure 1) does suggest some excess emission at $\sim$6 keV, and -
most interestingly - shows evidence of absorption near $\sim$8 keV in both data sets. When extrapolated to 0.3 keV,
the 2--11 keV power law fits to both pn and MOS data reveal the previously reported strong `soft excess' (figure 2).

\begin{figure}                                                          
\centering                                                              
\includegraphics[width=6.3cm, angle=270]{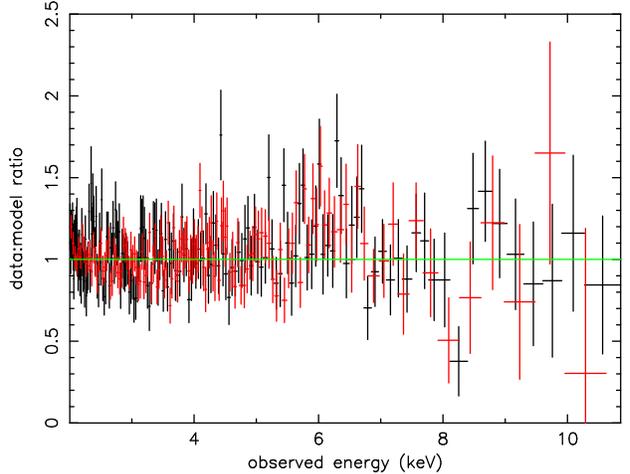}                     
\caption                                                                
{Ratio of the EPIC pn (black) and MOS (red) spectral data to a   
simple power law model fitted between 2-11 keV for \pg.  
The plot shows a broad excess near 6 keV and a narrow absorption feature at $\sim$8 keV.}      
\end{figure}                                                            
  
\begin{figure}                                                          
\centering                                                              
\includegraphics[width=6.3 cm, angle=270]{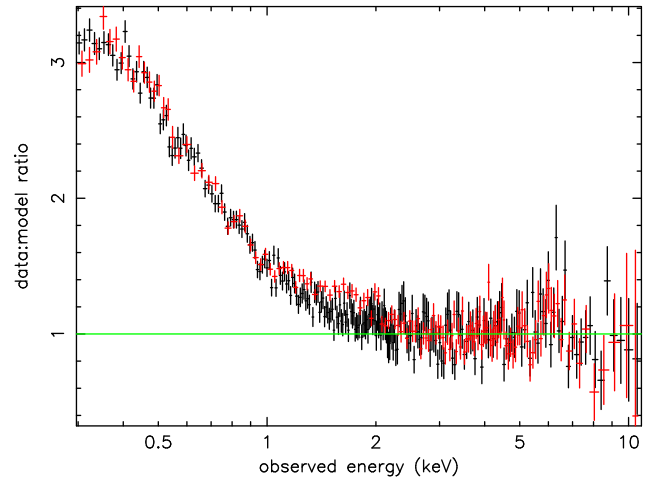}                     
\caption                                                                
{Extension to 0.3 keV of the 2-11 keV power law model fits for the pn (black) and MOS (red) spectral data, 
showing the strong soft excess in \pg.}
\end{figure}      

\subsection{Fe K emission and absorption features}

We then added further spectral components to match the features seen in the data, beginning with a 
gaussian emission line with energy, width and equivalent width (EW) as free parameters. This addition improved the
2--11 keV fit, to $\chi^{2}$/dof of 526/583, with a line energy (in the AGN rest frame) at 6.5$\pm$0.1 keV,
rms width $\sigma$ = 0.35$\pm$0.15 keV and EW = 0.25$\pm$0.11 keV. 

Next, we fitted the narrow absorption feature visible in figure 1 with a gaussian shaped absorption line, again with
energy, width and equivalent width free. The best-fit observed line energy was 8.18$\pm$0.10 keV, with an rms width of
$\leq$100 eV, and an EW of 170$\pm$60 eV. The addition of this gaussian absorption line improved the overall fit to
$\chi^{2}$/dof = 515/580, an improvement at 99.3 per cent confidence by the F-test (Bevington and Robinson 1992). We
then examined the EPIC data for other spectral features, excluding the region $\sim$1.8--2.3 keV where calibration
residuals associated with the Si K and Au M edges remain. The strongest feature is a possible absorption line at
$\sim$3.0 keV; fitting this with a gaussian line, again with energy and equivalent width free, further improved the 2--11 keV
fit,
to $\chi^{2}$/dof of 508/578  (98 per cent confidence). A note of caution is appropriate here, since the $\sim$3 keV feature
is only significant in the pn data. We reject a second feature of similar depth, apparent near 3.3 keV, where 
a gaussian
fit (figure 3) shows this `absorption line' to be marginally too narrow to match the EPIC resolution. 

Finally, we retain only the most convincing absorption features, at $\sim$8.18 keV and  $\sim$3.02 keV, in Table 1, 
which lists the
observed and AGN rest frame energy of each line, together with their most likely identifications and corresponding
outflow velocities. Since the Fe K-shell dominates X-ray absorption above $\sim$7 keV, with a series of resonance and
weaker satellite lines (eg Palmeri \et\ 2002) leading up to the absorption edges of He-like FeXXV at 8.76 keV and
H-like FeXXVI at 9.28 keV, the most likely interpretion of the $\sim$8.18 keV feature, where the narrow
profile indicates a line rather than an edge, is with the primary resonance absorption line in H- or He-like Fe.
These alternative identifications indicate
an outflow velocity of $\sim$0.22c or $\sim$0.26c, respectively. The absorption feature
at $\sim$3.02 keV has a most probable association with the primary resonance absorption line of He- or H-like S 
(as S K-shell absorption is likely to to be dominant in a highly ionised absorber in this energy range). We note the
implied outflow velocities from these alternatives are consistent with the values deduced for the ionised
Fe line, lending support to the overall interpretation. 

In summary, the detection of absorption lines in the EPIC spectra of \pg\ provides intriguing evidence of an ionised outflow, with a
velocity,
depending on the line identifications, in the range $\sim$0.20--0.26c. To attempt to reduce this uncertainty, and
find supporting evidence for this conclusion,
we
extend our search in Section 5 to the simultaneous RGS observation of \pg.

\begin{table*}
\centering
\caption{Candidate line identifications and corresponding outflow velocities in the parametric fit to the EPIC 
spectrum of \pg. Line energies 
are in keV.}

\begin{tabular}{@{}lccccc@{}}
\hline
Line  & $E_{obs}$ & $E_{source}$ & $E_{lab}$ &  outflow velocity  & EW (eV) \\

\hline

FeXXVI Ly$\alpha$ & 8.18$\pm$0.1 & 8.70 & 6.96 &  $\sim$0.22c & 170$\pm$60  \\
FeXXV 1s-2p & 8.18$\pm$0.1 & 8.70 & 6.70 &  $\sim$0.26c & 170$\pm$60  \\
SXVI Ly$\alpha$ & 3.02$\pm$0.1 & 3.21 & 2.62 &  $\sim$0.20c & 35$\pm$16 \\
SXV 1s-2p & 3.02$\pm$0.1 & 3.21 & 2.46 &  $\sim$0.26c & 35$\pm$16 \\

\hline
\end{tabular}
\end{table*}

\begin{figure}  \centering  \includegraphics[width=6.3 cm, angle=270]{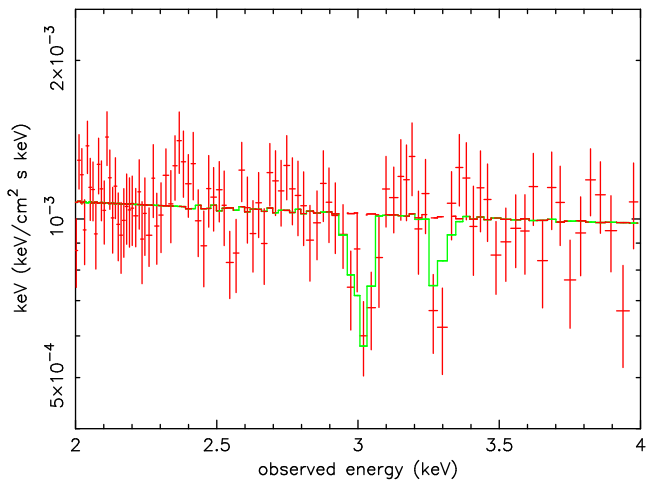}  \caption {Fit to spectral features
in the pn data between 2--5 keV. Possible identifications with S K lines are detailed in Table 1.}   \end{figure}

\section{Soft Excess}

Extending the 2--11 keV power law fits for both pn and MOS spectral data to 0.3 keV shows very clearly (figure 2) the
strong soft excess indicated in earlier observations. In the EPIC data this can be adequately  modelled with the
addition of 3 black body emission components, at kT=$\sim$70 eV, $\sim$130 eV and $\sim$330 eV, fitting the `gradual
soft excess' seen in \pg\ and which is apparently typical of higher luminosity AGN (Pounds and Reeves 2002). Based on
this fit we obtain an  average 0.3--10 keV flux for \pg\ of $1.1\times10^{-11}$~erg s$^{-1}$ cm$^{-2}$, corresponding
to a luminosity of $\sim 10^{44}$~erg s$^{-1}$ ($ H_0 = 75 $~km\,s$^{-1}$\,Mpc$^{-1}$). The 2--10 keV flux was
$3.8\times10^{-12}$~erg s$^{-1}$ cm$^{-2}$, with a corresponding luminosity of  $3\times 10^{43}$~erg s$^{-1}$.
Compared with the \xmm\ observation of PG1211+143, the soft excess in \pg\ is found to be 
significantly `hotter', as is evident from a comparison of figure 2 here with the similar plot for PG1211+143 (figure 3 in 
Pounds \et\ 2003).

\section{Spectral lines in the RGS data}

In Section 3.2 absorption lines in the EPIC spectrum have been  identified alternatively with resonance transitions in
He-like FeXXV  and SXV or H-like FeXXVI and SXVI.  The ambiguity in these identifications leaves the deduced outflow
velocity uncertain within the range $\sim$0.20-0.26c. To attempt to clarify the situation we then examined the simultaneous
\xmm\ grating spectra. We began by jointly  fitting the RGS-1 and RGS-2 data with a power law and black body continuum (from
the EPIC 0.3--11 keV fit) and examining the data:model residuals by eye. Only a few, weak spectral features were found, with
the most significant at the shortest wavelengths (consistent with a high ionisation parameter). Figure 4 reproduces a
section of the combined RGS data in the 8--15 Angstrom band, showing narrow absorption lines at $\sim$10.85 Angstrom and
$\sim$9.18 Angstrom, and a resolved emission line at $\sim$12.53 Angstrom. The observed wavelength and EW of each line was
determined by adding individual gaussian lines to the powerlaw plus blackbody continuum fit from EPIC. The statistical
quality of the fit was improved by the addition of the 3 lines, from $\chi^{2}$/dof = 90/102 to $\chi^{2}$/dof = 68/93 (99.8
percent confidence by the F-test). To attempt an initial identification of the absorption lines we noted that He- and H-like
Ne (and possibly several Fe L lines) dominate photoionised spectra at 8--15 Angstroms. Table 2 lists the line energies, for
the primary resonance line of NeX and NeIX, and it is encouraging that the pair of absorption lines are separated by the
correct wavelength ratio to fit the first 2 lines of either the NeX Lyman series or the 1s-2p and 1s-3p transitions of NeIX.
While this adds strength to the evidence for a high velocity outflow in \pg, the ambiguity relating to the He- or H-like
identifications remains. However, the apparent absence of Fe L absorption suggests a high ionisation parameter, while a fit
to the He-like lines would leave the absence of the NeX Ly$\alpha$ line (which should then appear at $\sim$10 Angstrom) a
problem.  We provisionally conclude, subject to testing with the XSTAR modelling in Section 6, that the H-like line
identifications give a physically more compatible description of the ionised outflow in \pg. Comparison of Tables 1 and 2
then require an outflow velocity in the range $\sim$0.17--0.22c, with an indication of a positive correlation of velocity
with ionisation energy. That interpretation would then suggest the emission observed at $\sim$12.53 Angstrom, and resolved
by the RGS, is primarily due to NeX Ly$\alpha$, corresponding to emission from matter moving at a mean velocity (to the line
of sight) of $\sim$0.03c. The strength and relatively low `blueshift' of this component indicates the outflow has a wide
cone angle, a point taken up again in Section 7.1, where we note that a broad outflow can be important in limiting sideways
leakage of the radiation flux responsible for accelerating the outflow.

As in our observation of PG 1211+143, the absorption line profiles are unresolved, again suggesting the absorbing
material is streaming outward with relatively little turbulence, and that we are viewing down (rather than across) the
flow. The narrow line widths also suggest the high column density inferred from the XSTAR model fitting (see Section 6)
may be an underestimate since no account is taken of saturation in the narrow line cores. Indeed, the
comparable strengths of the NeX Ly$\alpha$ and $\beta$ absorption lines suggest the former is
strongly saturated.

\begin{table*}
\centering
\caption{Ne K emission and absorption lines identified in the RGS spectrum of \pg. All wavelengths in Angstroms}.

\begin{tabular}{@{}lccccccc@{}}
\hline
Line & $\lambda$$_{obs}$ & $\lambda$$_{source}$ & $\lambda$$_{lab}$ &  outflow velocity & EW (mA)\\

\hline

NeX L-alpha & 12.53 $\pm$0.1 & 11.78 & 12.13 & $\sim$0.03c & 125$\pm$30  \\
NeIX-f 1s-2p & 12.53 $\pm$0.1 & 11.78 & 13.65 & $\sim$0.15c & 125$\pm$30 \\
NeX L-alpha & 10.85 $\pm$0.1 & 10.20 & 12.13 &$\sim$0.17c & 110$\pm$35  \\
NeIX 1s-2p & 10.85 $\pm$0.1 & 10.20 & 13.45 & $\sim$0.27c & 110$\pm$35  \\
NeX L-beta & 9.18 $\pm$0.1 & 8.63 & 10.24 & $\sim$0.17c & 90$\pm$45 \\
NeIX 1s-3p & 9.18 $\pm$0.1 & 8.63 & 11.55 & $\sim$0.28c & 90$\pm$45 \\

\hline
\end{tabular}
\end{table*}

\begin{figure} 
\centering 
\includegraphics[width=6.3 cm, angle=270]{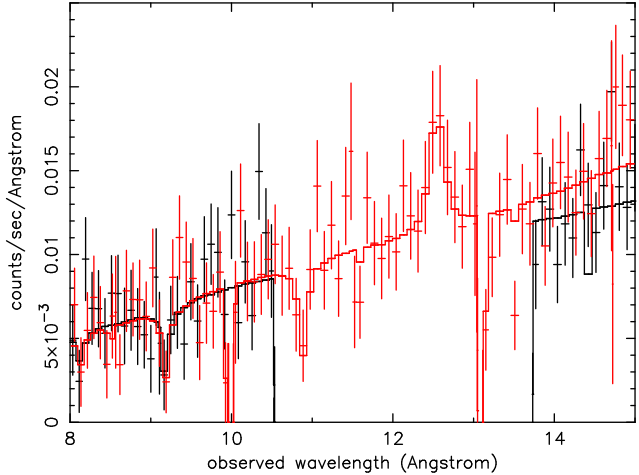} 
\caption {Fit to spectral features in the RGS data between 8--14 Angstrom. Possible identifications with Ne K lines 
are detailed
in Table 2. Data dropouts near 10, 10.5, 13.1 and 13.7 Angstrom are due to chip gaps in the RGS CCDs.}  
\end{figure}

\section{An ionised absorber model}

To quantify the highly ionised matter responsible for the observed absorption features, and check for physical consistency
of the candidate line identifications, we replaced the gaussian absorption lines in the above fits with a model comprising a
grid of photoionised absorbers based on the XSTAR code. These model absorbers cover a wide range of column density and
ionisation parameter $\xi$ (= $L/nr^2$), with outflow (or inflow) velocities as a variable parameter. All abundant elements from C to Fe are
included with the relative abundances as a variable input parameter. To limit processing time the XSTAR models assume a
fixed width for each absorption line of 1000 km s$^{-1}$ FWHM. Given the paucity of data from the \pg\ spectral analysis we
included only one absorbing column in the fit, assumed to fully cover the underlying continuum, recognising this would 
probably `average out' the optimal parameters for ions as disparate as FeXXVI and NeX. Assuming solar abundances, a best-fit
ionisation parameter of log$\xi$ of 3.7$\pm$0.2, with a column density of $N_{H}=4\times10^{23}\rm{cm}^{-2}$, was
found to reproduce the observed absorption line strengths of FeXXVI, SXVI and NeX, with a nett blueshift of 0.15 (or
$\sim$0.21c, in the AGN rest frame).  Apart from a minor contribution from FeXXV, the main absorption lines produced by this
highly ionised column were all of hydrogenic ions. An alternative fit, at an ionisation parameter of log$\xi$ = 
3.2$\pm$0.15, was obtained by assuming line identifications with He-like Fe, S and Ne. However, at this lower ionisation
level, significant Fe L absorption lines are predicted in the 10-15 Angstrom band, which are not seen. We therefore
conclude that the XSTAR fitting favours the high ionisation parameter solution, and an outflow velocity of $\sim$0.2c.

\section{Discussion}

The \xmm\ observation of \pg\ has revealed an X-ray spectrum with intriguingly similar (albeit weaker) features to those
recently reported for another bright, narrow line quasar PG1211+143 (Pounds \et\ 2003).  The soft X-ray excess reported in
previous observations is confirmed, being somewhat `hotter' and less strong than for PG1211+143. Fe K emission is again
detected and modelled by a broad, but non-relativistic gaussian line. Of most interest is finding further evidence of an
absorption line structure, in both  EPIC and RGS data, indicating a high column, high ionisation absorber, outflowing at an
even higher velocity (than found for PG1211+143) of $\sim$0.2c. We attribute the relatively faint absorption lines to the
higher ionisation (lower opacity) of the outflow, despite the best-fit column density again being remarkably high. As for
PG1211+143 we note that most of the uncertainty in the derived column densities is on the upside, since partial covering and
accounting for saturation in the  relatively narrow line profiles would both increase the above values. 

\pg\ adds to a growing list of AGN showing X-ray evidence for high velocity ionised outflows, also including PG1211+143
(Pounds \et 2003), the ultra-luminous quasar PDS456 (Reeves \et\ 2003) and the BAL quasar, APM 08279+5255, reported to have
strongly blue-shifted resonance absorption lines of Fe XXV or XXVI by Chartas \et\ (2002). The most important implication of
finding such absorption in the hard X-ray band is that the required column densities are higher (cf with those seen in the
UV), and the mass and kinetic energy in the outflows are correspondingly more significant. It appears that highly ionised,
high velocity gas, capable of imparting Fe K absorption features, may be a major component of (at least some) AGN that has
remained undetected prior to the improved sensitivity of observations in the $\sim$7--10 keV band.  

\subsection{A massive ionised outflow} As in the case of PG1211+143, the high column density of ionised matter in the
line-of-sight to \pg\ implies - for the simplest assumption of a radial outflow - that the flow will be optically thick at
small radii. We show below that the hot `photosphere' provides a major part of
the energetically dominant thermal continuum (BBB) emission for \pg. Other important implications of the observed high velocity
outflow, which we assess below in terms of a
physical model of winds from black holes accreting at or above the Eddington limit (King and Pounds 2003; hereafter
KP03), are a significant mass loss and substantial kinetic energy associated with the outflow.

As shown in the analysis of PG1211+143 (Pounds \et\ 2003), for a radial outflow with constant (coasting) speed $v$,
occupying a solid angle of $4\pi b$ steradians, mass conservation implies the outflow is optically thick at a radius $R_{\rm
ph}$, where  \begin{equation} {R_{\rm ph}\over R_{\rm s}} = {1\over 2\eta b}{c\over v}{\dot M_{\rm out}\over \dot M_{\rm
Edd}} \label{1} \end{equation}  Here $\dot M_{\rm out}$ is the outflow mass rate and $\dot M_{\rm Edd}$ the Eddington
accretion rate. $R_{\rm s} = 2GM/c^2$ is the Schwarzschild radius for mass $M$ and $\eta \sim 0.1$ is the assumed accretion
efficiency. Since the outflow is optically thick, most of the photons have scattered and given up at least their original
momentum. This will be assured if photon leakage from the sides of the outflow cone is small, ie the collimation of the
outflow is not too extreme. 

As in KP03 we assume that the outflow is launched from the photosphere at the local escape velocity, ie
\begin{equation}
R_{\rm ph}\simeq R_{\rm esc} = {c^2\over v^2}R_{\rm s}.
\label{esc}
\end{equation}

With $\dot M_{\rm Edd}$=0.7$\Msun$ yr$^{-1}$, appropriate to a black hole mass of $M \sim 3 \times 10^{7}\Msun$, we then
find $\dot M_{\rm out}$=0.7b$\Msun$ yr$^{-1}$. Assuming the measured outflow velocity is the same as the launch velocity, the
associated kinetic energy is 8b$\times$$10^{44}$~erg s$^{-1}$.

Since the outflow is optically thick for $\mo \sim \me$,  much of the accretion luminosity generated  deep in the potential
well near $\rs$ must emerge as blackbody--like  emission from it. The quasi--spherical radiating area is

\begin{equation} A_{\rm phot} = 4\pi b\rp^2  
\end{equation} 
with an effective blackbody temperature (KP03, equation 18) of
\begin{equation} T_{\rm eff} = 10^5b^{1/4}\dot M_1^{-1}M_8^{3/4}~{\rm K}. \label{teff} 
\end{equation} 
where $\dot M_1 =
\mo/(1\msun~{\rm yr}^{-1}), M_8 = M/10^8\msun$. 

Assuming b$\sim$1 in \pg\ this gives  $T_{\rm eff} = 6\times 10^4$K, consistent with the optically thick flow being
a major component of the BBB (and bolometric luminosity) of \pg.

Confirmation of the physical consistency of the above analysis can be made by noting that, for a Compton thick wind driven by
radiation pressure, the momentum in the outflow must be of the same order as that in the Eddington-limited radiation field (KP03,
equation 12), as is indeed the case for \pg.   

Furthermore, the kinetic energy in the outflow should be lower than that
of the radiation field by a factor of order $v/c$ (equation 13 in KP03),
a requirement which again is met by the outflow parameters of \pg.

In summary, we conclude that the radiation driven, optically thick wind described in KP03 represents a
physically consistent model by which to assess the high velocity, highly ionised outflows detected in \pg\ and several other AGN.
A question remains as to how common are these phenomena, and - in turn - how common is supercritical accretion?

\subsection{ How common are massive outflows in AGN?}

As noted earlier the bolometric luminosity of \pg\ is of order $3\times$$ 10^{45}$~erg s$^{-1}$, which for a  reverberation
mass of $M \sim 3 \times 10^{7}\Msun$ (Kaspi \et\ 2000) implies accretion at close to the Eddington rate. PG1211+143 has a
very similar luminosity and black hole mass (Kaspi \et\ 2000) and - taken together - the  \xmm\ observations suggest that  a
high velocity, highly ionised outflow may be a new signature of a high (super-Eddington?) accretion rate in AGN. A strong
soft X-ray excess (seen in earlier data as a steep soft X-ray spectrum) then follows naturally from the physical model
sketched in Section 7.1. It is interesting to note that the detailed differences  in the observed properties of \pg\ and
PG1211+143 are also broadly consistent with the physical model described in KP03.  In particular, we associate the more
highly ionised/higher velocity outflow for \pg\ with a smaller launch radius.  The smaller and hotter photosphere matches in
a general way the lower UV flux (m$\sim$14 compared m$\sim$13.3 from the simultaneous OM data), but `hotter' soft X-ray
excess in \pg\ compared with PG1211+143 (compare figure 2 in this paper with figure 3 in Pounds \et\ 2003).  Another
interesting difference evident in the simple power law fits to the EPIC data of \pg\ and PG 1211+143 is the absence  of an
extreme broad Fe K emission line in the former case. While, if truly a relativistic Fe K line, this could relate to a
difference in reflection from  the innermost accretion disc, a simpler alternative is that the partial covering by a column
of moderately ionised gas (proposed to explain the extreme Fe K `line' in PG1211+143) is simply absent from the more highly
ionised flow in \pg.    In summary, the \xmm\ observations of \pg\ and PG1211+143 provide strong evidence of a massive
outflow of ionised gas, which may well be a characteristic of AGN accreting at or above the Eddington limit. The question of
how common are these outflows might then be transposed to the more fundamental question of how common is Eddington or
super-Eddington accretion in AGN?

Until recently the evidence for high velocity outflows in AGN has come from UV studies of BAL  quasars, where a variety of,
mainly high-ionisation, UV resonance transitions exhibit velocity widths up to $\sim 30000$ km s$^{-1}$ (e.g. Weymann et al.
1991). The BAL property, involving some 10\% of optically selected  quasars, is usually interpreted as viewing the central
continuum source along a particular line-of-sight  tangential to the accretion disc. At X-ray energies most BAL quasars are
`x-ray weak', a property explained by line-of-sight absorption in moderately ionised gas with column densities up to
$N_{H}=10^{23}\rm{cm}^{-2}$ (e.g. Hamann 1998; Sabra \& Hamann 2001; Gallagher \et\ 2002). Crucially, these observations
have not revealed whether the X-ray absorbing column is moving or stationary. The extended \chandra\ observation of the
gravitationally lensed high redshift BAL quasar APM 08279+5255 (Chartas \et\ 2002) was an exception, showing  components of
the absorbing column to be highly ionised and outflowing at $\sim$0.2c and $\sim$0.4c. Taken together with the evidence
reported here (and for PG1211+143 and PDS456) it seems a reasonable speculation that a high column density and high velocity
(hence massive) outflow may be a common property of luminous AGN. This invites the further intriguing speculation that
accretion at the Eddington rate is equally common, extending the suggestion by Becker \et\ (2000) who proposed that BAL
objects may be young or have recently been fuelled.

Observationally, the detection of highly ionised outflows remains challenging since the gas is largely transparent, showing
its most unambiguous features in the X-ray band above $\sim$7 keV where current in-orbit instruments have low sensitivity.
Nevertheless, future longer exposures with \xmm\ and \chandra\ should provide a better assessment of this important new
component of (at least high accretion rate) AGN. 

\section{Conclusions}

(1) Archival \xmm\ data of the bright quasar \pg\ has revealed further evidence of a high velocity ionised outflow
in a PG quasar.

(2) An implication of the observed high column density is that the inner flow will be optically thick, providing a
natural explanation for the dominant BBB (and strong soft X-ray emission) in \pg.

(3) Efficient acceleration requires the outflow in \pg\ to be only weakly collimated, which then leads to estimates of the
mass and kinetic energy of the flow representing a significant fraction of the accretion mass rate and energy budget. 

(4) We suggest the above properties may be common in AGN accreting at or above the Eddington limit.  

\section*{ Acknowledgements }
The results reported here are based on observations obtained with \xmm, an ESA science mission with instruments and
contributions directly funded by ESA Member States and the USA (NASA). The authors wish to thank the SOC and SSC teams
for organising the \xmm\ observations and initial data reduction. ARK gratefully acknowledges a Royal Society Wolfson
Research Merit Award.

\end{document}